\documentclass[runningheads,a4paper]{llncs}
  \usepackage{mathptmx}
  
\usepackage{amssymb}
\usepackage{graphicx}
\usepackage{color}
\usepackage{enumerate}
\usepackage{calc}
\usepackage{multirow}
\usepackage{url}

	\newcommand{\ccluster}{\texttt{Ccluster}}
	\newcommand{\mpsolve}{\texttt{MPSolve}}
	\newcommand{\fsolve}{\texttt{fsolve}}
	\newcommand{\unisolve}{\texttt{unisolve}}
	\newcommand{\secsolve}{\texttt{secsolve}}
	\newcommand{\eigensolve}{\texttt{eigensolve}}
	
	\newcommand{\intcompare}{\texttt{IntCompare}}
	\newcommand{\wt}[1]{\widetilde{#1}}

	\newcommand{\wtTGk}{\wt{T}^G_k}
	\newcommand{\wtTGo}{\wt{T}^G_0}
	
	\newcommand{\wtTGs}{\wt{T}^G_\ast}
	
	\newcommand{\true}{\mbox{\bf true}}	
	\newcommand{\false}{\mbox{\bf false}}	
	\newcommand{\vareps}{\varepsilon}
	\newcommand{\maple}{\mbox{\texttt{Maple}}}
	\newcommand{\dt}[1]{\textbf{#1}}
	\newcommand{\CC}{{\mathbb C}}
	\newcommand{\ib}{\subseteq }
	\newcommand{\set}[1]{\left\{ #1 \right\}}
	\newcommand{\dd}{ ,\ldots , }
	\newcommand{\bitem}{\begin{itemize}}
	\newcommand{\eitem}{\end{itemize}}
	\newcommand{\benum}{\begin{enumerate}}
	\newcommand{\eenum}{\end{enumerate}}
	\newcommand{\wtO}{\widetilde{O}}
	\newcommand{\as}{\mathrel{:=}}
	\newcommand{\refFig}[1]{Figure~\protect{\ref{fig:#1}}}
	\newcommand{\refTab}[1]{Table~\protect{\ref{tab:#1}}}
	\newcommand{\cored}[1]{\textcolor{red}{#1}}
	\newcommand{\cocyan}[1]{\textcolor{cyan}{#1}}
	\newcommand{\ttt}[1]{{\tt #1}}
	\newcommand{\ii}{\mbox{\bf i}}
	\newcommand{\mathematica}{\mbox{\texttt{Mathematica}}}
	\newcommand{\ass}{\leftarrow}
	\newcommand{\ceil}[1]{\left\lceil {#1} \right\rceil}

	\newenvironment{boxp}[2][0.65]{\begin{center}
		\fbox{\begin{minipage}{#1 \textwidth}
		#2
		\end{minipage}}
		\end{center}}{
		}
	\newenvironment{prog}{\begin{tabbing}
	 xxxx\=xxxx\=xxxx\=xxxx\=xxxx\=xxxx\=xxxx\=xxxx\=xxxx\=xxxx\=xxxx\=xxxx\=xxxx\=
	\kill\\}{
		\end{tabbing}}
	\newenvironment{progb}[2][2]{ 
		\begin{center}
		  \fbox{\begin{minipage}{0.75\textwidth}
	        	\vspace*{-#1\abovedisplayskip}
			\begin{prog}#2\end{prog}
	          \end{minipage}}
	        \end{center}
		}{}
	\newcounter{indentcounter1}
	\newcounter{indentcounter2}
	\newcounter{indentcounter3}
	\newcommand{\Indent}[1][5]{
		\setcounter{indentcounter1}{#1}		
		\setcounter{indentcounter2}{1}		
		\setcounter{indentcounter3}{
		    \value{indentcounter1}*\value{indentcounter2}}
		\hspace*{\value{indentcounter3} mm}}
	\newcommand{\lline}[1][0]{
	       \ \\ \Indent[#1]\texttt {}
	}
	\newcommand{\Comment}[1]{\quad\cocyan{\mbox{$\triangleleft\;\;$}{\em #1}} }
	\newcommand{\OUTPUt}{\mbox{\textsf{Output}}}

	\newcommand{\unresolved}{\mbox{\bf unresolved}}
	\newcommand{\Return}{\mbox{\bf Return}}
	\newcommand{\clus}{\mbox{\tt \#Clus}}
	\newcommand{\sols}{\mbox{\tt \#Sols}}
	\newcommand{\Ber}{\mbox{\tt Bern}}	
	\newcommand{\Mig}{\mbox{\tt Mign}}	
	\newcommand{\Wil}{\mbox{\tt Wilk}}	
	\newcommand{\Spi}{\mbox{\tt Spir}}	
	\newcommand{\WilM}{\mbox{\tt WilkMul}}	
	\newcommand{\MigC}{\mbox{\tt MignClu}}	
	\newcommand{\NesC}{\mbox{\tt NestClu}}	

	\newcommand{\cheeOK}[2]{ #2 }
	\newcommand{\remiOK}[2]{ #2 }
        \newcommand{\panOK}[2]{ #2 }

\begin{document}
\mainmatter
\title{Implementation of a Near-Optimal Complex Root Clustering
	Algorithm}
\titlerunning{Implementation of Complex Root Clustering}
\author{R\'emi Imbach \inst{1}
		\thanks{
		R\'emi's work has received funding from the European
		Union's Horizon 2020 research and innovation programme under
		grant agreement No. 676541.
		}
	\and Victor Y. Pan\inst{2}
		\thanks{
	    Victor's work is supported by NSF Grants
	    \#~CCF-1116736 and \#~CCF-1563942 and by PSC CUNY Award
		698130048.}
	\and Chee Yap \inst{3}
		\thanks{
		    Chee's work is supported by
			NSF Grants \#~CCF-1423228 and \#~CCF-1564132.}
	    }
\authorrunning{Imbach-Pan-Yap}
\institute{
  TU Kaiserslautern\\
  Email: \email{imbach@mathematik.uni-kl.de}\\
  \url{www.mathematik.uni-kl.de/en/agag/members/}
\and
  City University of New York\\
  Email: \email{victor.pan@lehman.cuny.edu}\\
  \url{http://comet.lehman.cuny.edu/vpan/}
\and
  Courant Institute of Mathematical Sciences\\
  New York University, USA\\
  Email: \email{yap@cs.nyu.edu}\\
  \url{www.cs.nyu.edu/yap/}
}

\maketitle

\begin{abstract}
    We describe \ccluster, 
    a software for computing natural $\vareps$-clusters of complex roots
    in a given box of the complex plane. This algorithm from
    Becker et al.~(2016) is near-optimal when applied to
    the benchmark problem of isolating
    all complex roots of an integer polynomial.
    It is one\footnote{
	    Irina Voiculescu informed
	    us that her student Dan-Andrei Gheorghe has independently
	    implemented the same algorithm
	    in a Masters Thesis Project (May 18, 2017) at Oxford University.
	    Sewon Park and Martin Ziegler at KAIST, Korea, 
	    have implemented a modified version of Becker et al.~(2016)
	    for polynomials having only real roots
	    being the eigenvalues of 
	    symmetric square matrices with real coefficients.
	    See the technical report CS-TR-2018-415 at 
	    \url{https://cs.kaist.ac.kr/research/techReport}.
	    }
    of the first implementations of a near-optimal algorithm
    for complex roots.
    We describe some low level techniques for speeding up the algorithm. 
    Its performance is compared with the well-known \mpsolve\ library
    and \maple.

\end{abstract}
\section{Introduction}
\label{sec_intro}
	The problem of root finding for a polynomial $f(z)$ is 
	a classical problem from antiquity, but remains the
	subject of active research to the present
	\cite{emiris-pan-tsigaridas:handbk:14}.
	We consider a classic version of root finding:
	\begin{boxp}[0.8]{
	\dt{Local root isolation problem}:\\\em
	    {\bf Given:} a polynomial $f(z)\in\CC[z]$,
	    	a box $B_0\ib\CC$, $\vareps>0$.\\
	    {\bf Output:} a set $\set{\Delta_1\dd \Delta_k}$
		of pairwise-disjoint discs of radius $\le\vareps$,
		each containing a unique root of $f(x)$ in $B_0$.
	}
	\end{boxp}
	It is local because we only look for roots in a locality, as
	specified by $B_0$. 
	The local problem is useful in applications
	(especially in geometric computation) where we
	know where to look for the roots of interest.
	There are several variants of this problem:
	in the \dt{global version},
	we are not given $B_0$, signifying that we
	wish to find all the roots of $f$.  The global version
	is easily reduced to the local one by specifying a
	$B_0$ that contains all roots of $f$.
	If we omit $\vareps$, it amounts to setting $\vareps=\infty$,
	representing the pure isolation problem.

	Our main interest is a
	generalization of root isolation, to the lesser-studied
	problem of root clustering
	\cite{hribernig-stetter:cluster-zeros:97,niu+2:zeros-analytic:07,giusti+2:zeros-analytic:05}.
	It is convenient to introduce two definitions:
	for any set $S\ib\CC$, let $Z_f(S)$ denote the
	set of roots of $f$ in $S$, and let $\#_f(S)$
	count the total multiplicity of the roots in $Z_f(S)$.
	Typically, $S$ is a disc or a box.  For boxes and discs,
	we may write $kS$ (for any $k>0$) to denote the dilation
	of $S$ by factor $k$, keeping the same center.
	The following problem was introduced in  
	\cite{yap-sagraloff-sharma:cluster:13}:
	\begin{boxp}[0.8]{
	\dt{Local root clustering problem}:\\\em
	    {\bf Given:}
	    a polynomial $f(z)$, a box $B_0\ib\CC$, $\vareps>0$.\\
	    {\bf Output:}
	    a set of pairs $\set{(\Delta_1,m_1)\dd (\Delta_k,m_k)}$ 
		where
		\bitem
	    \item $\Delta_i$'s are pairwise-disjoint discs of radius
		$\le\vareps$,
	    \item $m_i=\#_f(\Delta_i)=\#_f(3\Delta_i)$
		for all $i$, and
	    \item
		$ Z_f(B_0) \ib \bigcup_{i=1}^k Z_f(\Delta_i).$
		\eitem
	}
	\end{boxp}
	This generalization of root isolation
	is necessary when we consider polynomials
	whose coefficients are non-algebraic
	(or when $f(z)$ is an analytic function, 
	as in \cite{yap-sagraloff-sharma:cluster:13}).
	The requirement that $\#_f(\Delta_i)=\#_f(3\Delta_i)$
	ensures that our output clusters are \dt{natural}
	\cite{becker+4:cluster:16};   
	a polynomial of degree $d$ has at most $2d-1$ natural clusters
	(see \cite[Lemma 1]{yap-sagraloff-sharma:cluster:13}).
	The local root clustering algorithm for analytic functions
	of \cite{yap-sagraloff-sharma:cluster:13}
	has termination proof, but no complexity analysis.
	By restricting $f(z)$ to a polymomial, 
	Becker et al.~\cite{becker+3:cisolate:18} succeeded in 
	giving an algorithm and also its complexity analysis
	based on the geometry of the roots.  When applied to
	the \dt{benchmark problem}, where $f(z)$ is an integer polynomial
	of degree $d$ with $L$-bit coefficients, the
	algorithm can isolate all the
	roots of $f(z)$ with bit complexity $\wtO(d^2(L+d))$.
	Pan \cite{pan:poly-roots:02} 		
	calls such bounds \dt{near-optimal} (at least when $L\ge d$).
	The clustering algorithm studied in this paper
	comes from \cite{becker+4:cluster:16}, which in turn
	is based on \cite{becker+3:cisolate:18}.
	Previously, the Pan-Sch\"onhage algorithm
	has achieved near-optimal bounds with
	divide-and-conquer methods \cite{pan:poly-roots:02},
	but \cite{becker+3:cisolate:18,becker+4:cluster:16}
	was the first {\em subdivision} algorithm to achieve the
	near-optimal bound for complex roots.
	For real roots, Sagraloff-Mehlhorn
	\cite{sagraloff-mehlhorn:real-roots:16}
	had earlier achieved near-optimal bound via subdivision.
	
	Why the emphasis on ``subdivision''?
	It is because such algorithms are implementable
	and quite practical (e.g., \cite{rouillier-zimmermann:roots:04}).
	Thus the near-optimal real subdivision algorithm of
	\cite{sagraloff-mehlhorn:real-roots:16}
	was implemented shortly after its discovery, and reported in
	\cite{kobel-rouillier-sagraloff:for-real:16}
	with excellent results.  In contrast, all the asymptotically
	efficient root algorithms (not necessarily near-optimal)
	based on divide-and-conquer methods
	of the last 30 years have never been implemented;
	a proof-of-concept implementation of Sch\"onhage's
	algorithm was reported in Gourdon's thesis
	\cite{gourdon:thesis:96}).
	Computer algebra systems mainly rely on algorithms with
	a priori guarantees of correctness.  But in practice,
	algorithms without such guarantees are widely used.
	For complex root isolation, one of
	the most highly regarded multiprecision software is \mpsolve\
	\cite{bini-fiorentino:design:00}.
	The original algorithm in \mpsolve\ was based
	on Erhlich-Aberth (EA) iteration; but since 2014, a
	``hybrid'' algorithm \cite{bini-robol:secular:14} was introduced.
	It is based on the secular equation,
	and combines ideas from EA and \eigensolve~
	\cite{fortune:eigensolve:02}.
	These algorithms are inherently global solvers
	(they must approximate {\em all} roots of a polynomial
	simultaneously).  Another theoretical limitation is that
	the global convergence of these methods is not proven.

	In this paper, we give a preliminary report
	about \ccluster, our implementation of the
	root clustering algorithm from \cite{becker+4:cluster:16}.
	
	\vspace*{-2em}
	\begin{figure}[!h]
	 \begin{minipage}{0.5\linewidth}
	 \centering
	  \includegraphics[width=5.5cm]{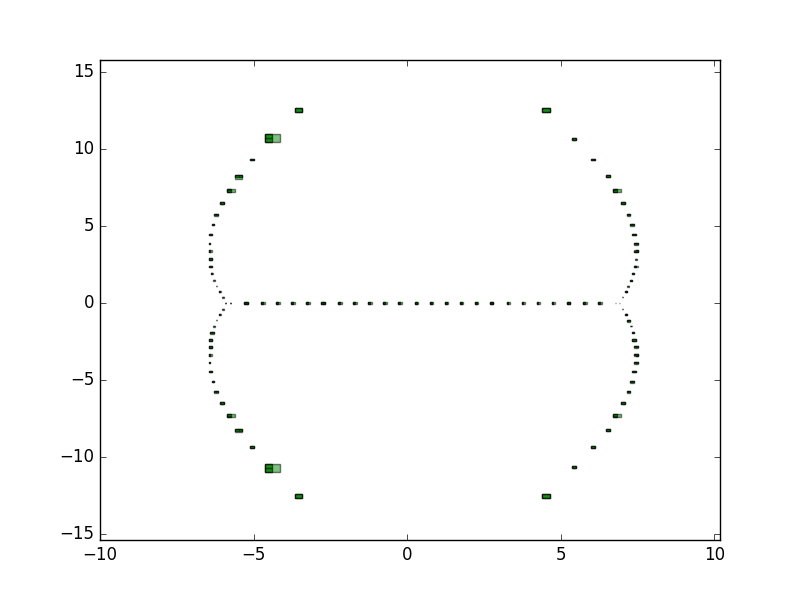}
	 \end{minipage}
	 \begin{minipage}{0.5\linewidth}
	 \centering
	  \includegraphics[width=5.5cm]{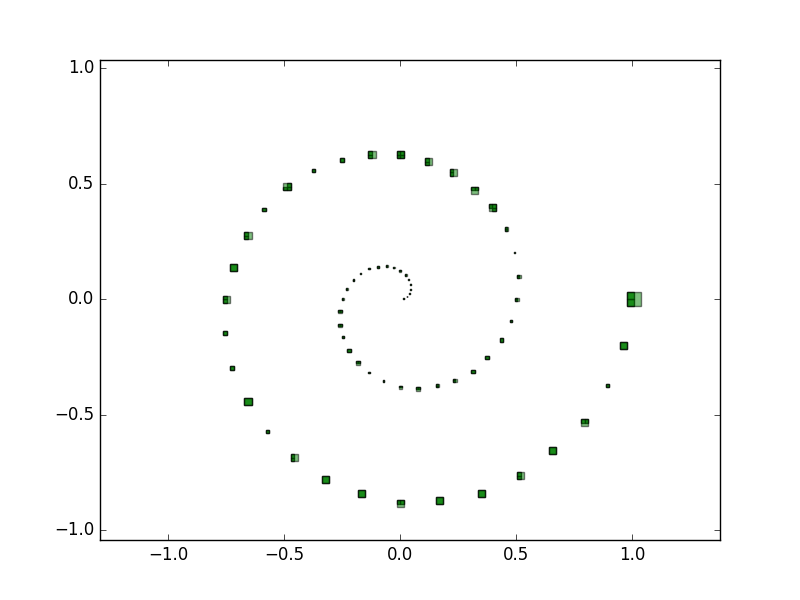} 
	 \end{minipage}
	    \caption{
	    {\bf Left:} the connected components isolating all roots
	    of the Bernoulli polynomial of degree 100.
	    {\bf Right:} the connected components isolating all roots of the
		  Spiral polynomial of degree 64. 
		  }
	 \label{fig:bernoulli_sols}
	\vspace*{-1em}
	\end{figure}

	To illustrate the performance for the local versus global
	problem, consider the Bernoulli polynomials
	$\Ber_d(z) \as \sum_{k=0}^{d} {{d}\choose{k}}b_{d-k}z^k$ where
	$b_i$'s are the Bernoulli numbers.
	\refFig{bernoulli_sols}(Left) shows the
		graphical output of \ccluster\ for $\Ber_{100}(z)$.
	\refTab{bernoulli} has four timings $\tau_{X}$
	(for $X=\ell, g, u, s$)
		in seconds:  $\tau_\ell$ is the time for solving
		the local problem over a box $B_0=[-1,1]^2$;
		$\tau_g$ is the time for the global problem
		over the box $B_0=[-150,150]^2$ (which contains all the roots).
		The other two timings from \mpsolve\
		($\tau_u$ for unisolve, $\tau_s$ for secsolve)
		will be explained later.
	For each instance, we also indicate the numbers of
	solutions (\sols) and clusters (\clus).
	When \sols\ equals \clus, we know the roots are isolated.
	Subdivision algorithms like ours naturally solve the local
	problem, but \mpsolve\ can only solve the global problem.
	\refTab{bernoulli}
	shows that \mpsolve\ remains unchallenged for
	the global problem.  But in applications where locality can
	be exploited, local methods may win, as seen in the last two
	rows of the table. The corresponding time for \maple's \fsolve\
	is also given; \fsolve\ is not a guaranteed algorithm and may fail.

	\begin{table}
	\begin{center}
	{\scriptsize
	\begin{tabular}{l||c|c|c||c|c|c||c|c||c|}
		& \multicolumn{3}{|c||}{\ccluster~ local ($B_0=[-1,1]^2$)}
		& \multicolumn{3}{|c||}{\ccluster~ global ($B_0=[-150,150]^2$)}
			    & \texttt{unisolve} 
			    & \texttt{secsolve} 
			    & \texttt{fsolve} \\\hline
		d& (\sols:\clus) & (depth:size) & $\tau_\ell$ (s)
			    & (\sols:\clus) & (depth:size) & $\tau_g$ (s)
			    & $\tau_u$ (s) & $\tau_s$ (s) & $\tau_f$ (s) \\\hline\hline
		64&(4:4)&(9:164)&0.12
			    &(64:64)&(17:1948)&2.10
			    & 0.13& \cored{0.01} & 0.1\\\hline
		128&(4:4)&(9:164)&0.34
			    &(128:128)&(16:3868)&9.90
			    & 0.55& \cored{0.05} & 6.84\\\hline
		191&(5:5)&(9:196)&0.69
			    &(191:191)&(17:5436)&32.5
			    & 2.29& \cored{0.16}&50.0\\\hline
		256&(4:4)&(9:164)&0.96
			    &(256:256)&(17:7300)&60.6
			    & 3.80& \cored{0.37}&$>1000$\\\hline
		383&(5:5)&(9:196)&2.06
			    &(383:383)&(17:11188)&181
			    &$>1000$& \cored{1.17}&$>1000$\\\hline
		512&(4:4)&(9:164)&\cored{2.87}
			    &(512:512)&(16:14972)&456
			    &$>1000$& 3.63&$>1000$\\\hline
		767&(5:5)&(9:196)&\cored{6.09}
			    &(767:767)&(17:22332)&1413
			    &$>1000$& 10.38&$>1000$\\\hline
	\end{tabular}
	}
	    \caption{Bernoulli Polynomials with five timings:
	    local ($\tau_\ell$), global ($\tau_\ell$), unisolve ($\tau_\ell$),
	    secsolve ($\tau_\ell$)
	    and \maple's \fsolve ($\tau_f$).}
	    \label{tab:bernoulli}
	    \vspace*{-2em}
	\end{center}
	\end{table}

\remiOK{\subsection{Overview of Paper}}{\noindent\textbf{Overview of Paper}}
	In Section 2, we describe the experimental setup for \ccluster.
	Sections 3-5 describe some techniques for speeding up the basic
	algorithm.  We conclude with Section 6.

\section{Implementation and Experiments}
	The main implementation of \ccluster\ is in \texttt{C} language.
	We have an interface for 
	\texttt{Julia}\footnote{\url{https://julialang.org/}.
	Download our code in
	{\scriptsize\ttt{https://github.com/rimbach/Ccluster}}.
	    }.
	We based our big number computation on the
	\texttt{arb}\footnote{\url{http://arblib.org/}.
	Download our code in
	{\scriptsize\ttt{https://github.com/rimbach/Ccluster.jl}}.
	}
	library.
	The \texttt{arb} library implements ball arithmetic for real numbers, 
	complex numbers and polynomials with complex coefficients. Each
	arithmetic operation is carried out with error bounds.

	\paragraph{Test Suite}
	We consider 7 families of polynomials,
	classic ones as well as some new ones constructed
	to have interesting clustering or multiple root structure.
	\benum[(F1)]
    \item
	The Bernoulli polynomial 
	$\Ber_d(z)$ of degree $d$
	\remiOK{was described earlier in the introduction.}{is described in Section~\ref{sec_intro}.}
    \item
	The Mignotte polynomial $\Mig_d(z;a) \as z^d - 2(2^az-1)^2$
	for a positive integer $a$, 
	has two roots whose separation is near the
	theoretical minimum separation bound.
    \item
	The Wilkinson polynomials
	$\Wil_d(z)\as \prod_{k=1}^d (z-k)$.
    \item
	The Spiral Polynomial
		$\Spi_d(z) \as
		    \prod_{k=1}^d\Big(z - \frac{k}{d}e^{4k\ii \pi/n}\Big).$
		See 
		\refFig{bernoulli_sols}(Right) for $\Spi_{64}(z)$.
		\cheeOK{Since this definition involve the exponential
		function, how do you convert this into a polynonmial?
		}{}
		\remiOK{This is already a polynomial with coefficients in $\CC$.
		      The coefficients (that have the form $something\times
		      e^{something}$)
		      are approximated with respect to the working precision.
		      }{}
	\item 
	    Wilkinson Multiple:
	    $\WilM_{(D)}(z)\as \prod_{k=1}^D (z-k)^k$.
		$\WilM_{(D)}(z)$ has degree $d=D(D+1)/2$ where
		the root $z=k$ has multiplicity $k$ (for $k=1\dd D$).
	\item 
	    Mignotte Cluster: 
	    $\MigC_{d}(z;a,k) \as x^d - 2(2^az-1)^k(2^az+1)^k$.
	    This polynomial has degree $d$ (assuming $d\ge 2k$)
	    and has a cluster of $k$ roots near
	    $2^{-a}$ and a cluster of $k$ roots near $-2^{-a}$.
	    \cheeOK{Why can't we omit the factor of $2$
			and simply use $x^D - (2az-1)^k(2az+1)^k$?
		The same question for the $\Mig_d(z)$.}{}
	    \remiOK{For the Mignotte polynomial, I used more or less the
	    definition of Mignotte and Bugeaud in the first page of 
	          http://irma.math.unistra.fr/~bugeaud/travaux/PolSurvRev1.pdf
		  Notice that $a$ in the latter document is $2^a$ here, this is
		  to underline the importance of the bitsize of the parameter.
		  }{}

\begin{figure}[!h]
		 \begin{minipage}{0.5\linewidth}
		 \centering
		  \includegraphics[width=5.7cm]
		     {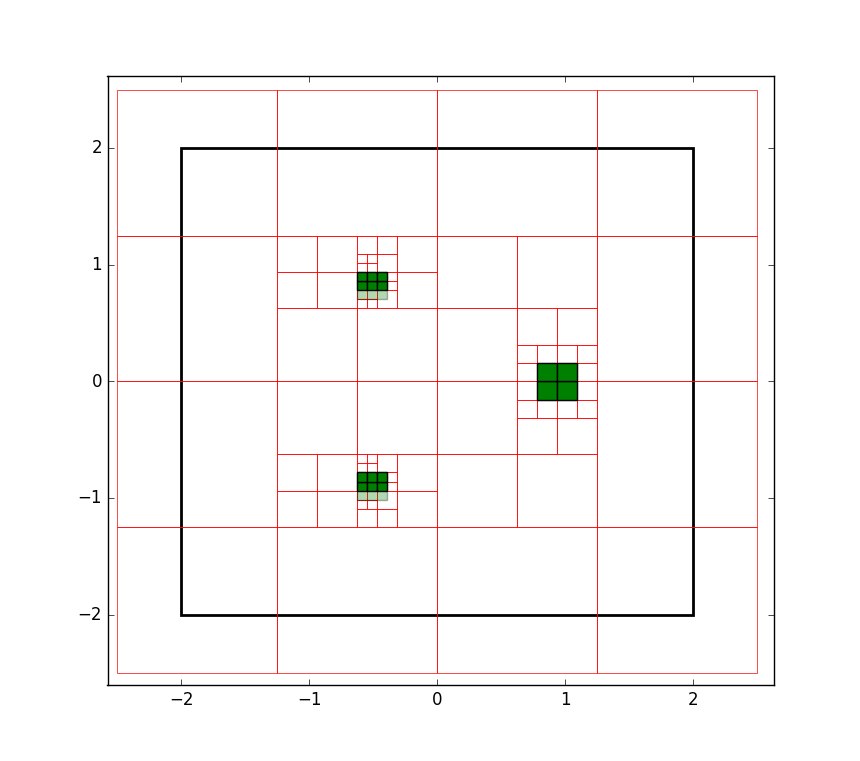}
		 \end{minipage}
		 \begin{minipage}{0.5\linewidth}
		 \centering
		  \includegraphics[width=4.56cm]
		      {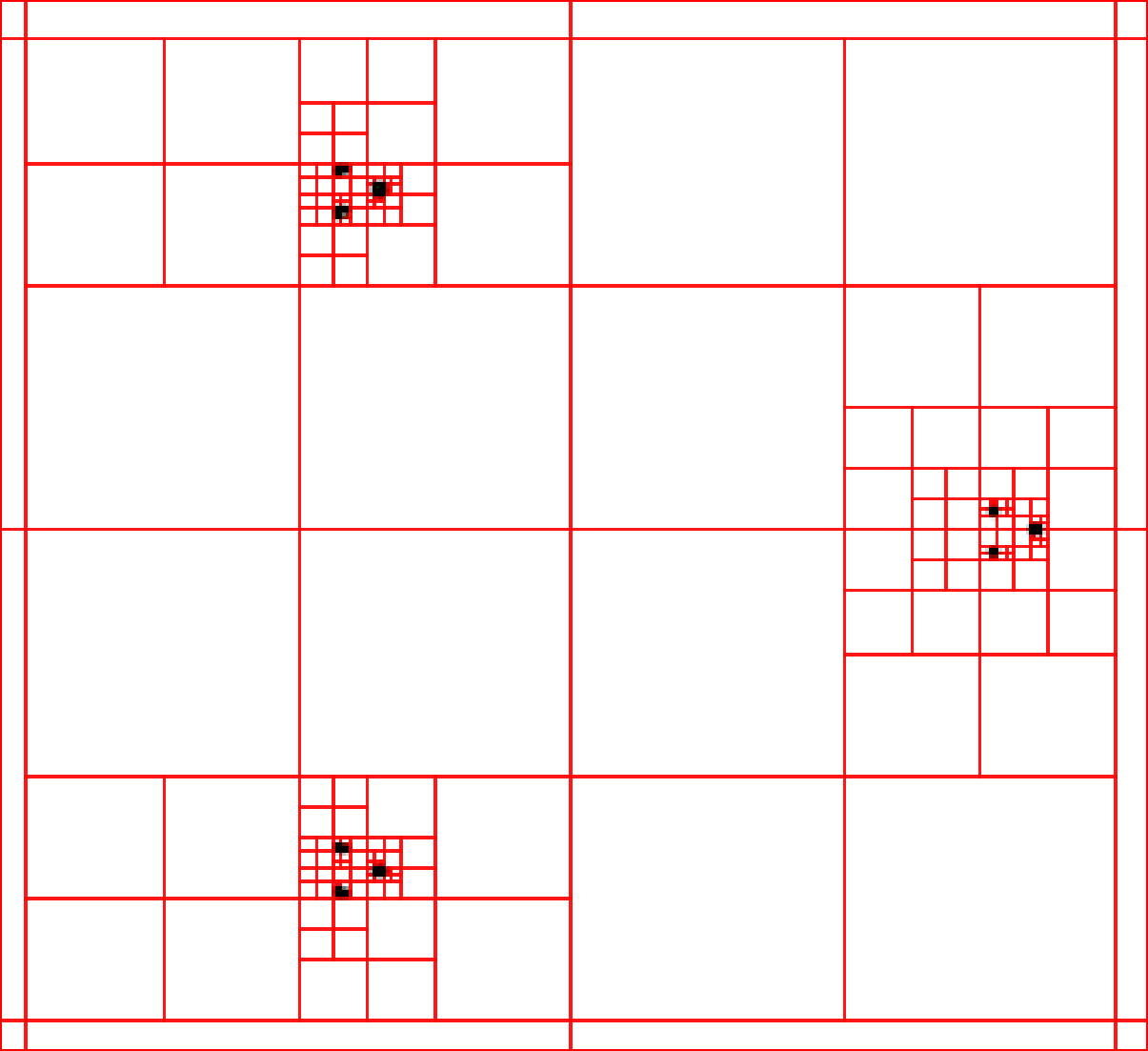}
		 \end{minipage}
		 \caption{{\bf Left:} 3 clusters of $\NesC_{(3)}$ found with
		 $\vareps=1$.  {\bf Right:} Zoomed view of 9 clusters of
		 $\NesC_{(3)}$ found with $\vareps=\frac{1}{10}$.
		 {\bf Note:} The initial box is
		 in thick lines; the thin lines show the subdivisions tree.
		 }
		 \label{fig:cluster_27_eps}
	\end{figure}

	\item 
	    Nested Cluster: $\NesC_{(D)}(z)$ has degree $d=3^D$
	    and is defined by induction on $D$:
		$\NesC_{(1)}(z)\as z^3-1$ with roots $\omega, \omega^2,
		\omega^3=1$ where $\omega=e^{2\pi\ii/3}$.
		Inductively, if the roots of $\NesC_{(D)}(z)$ are
		$\set{r_j: j=1\dd 3^D}$, then we define
			$\NesC_{(D+1)}(z) \as \prod_{j=1}^{3^D}
				\Big(z - r_j - \frac{\omega}{16^D}\Big)
				\Big(z - r_j - \frac{\omega^2}{16^D}\Big)
				\Big(z - r_j - \frac{1}{16^D}\Big)$
		See \refFig{cluster_27_eps} for the natural
		$\vareps$-clusters of $\NesC_{(3)}(z)$.
	\eenum

	\paragraph{Timing}
	Running times are sequential times
	on a Intel(R) Core(TM) i3 CPU 530 @ 2.93GHz machine with linux.
	\ccluster\ implements the algorithm
	described in \cite{becker+4:cluster:16} with differences
	coming from the improvements described in 
	Sections \ref{sec:pellet}-\ref{sec:escape} below.
	Unless explicitly specified,
	the value of $\epsilon$ for \ccluster\
	is set to $2^{-53}$; roughly speaking, it falls
	back to asking for 15 guaranteed decimal digits.
	
\paragraph{\mpsolve}
	\cheeOK{Should we compare with Maple also?  We might look more
	competitive this way.  Also, most people use maple!
	}{}
	For external comparison, we use \mpsolve. It was 
	shown 
	to be superior to major software such as \maple\
	or \mathematica\ \cite{bini-fiorentino:design:00}.
	There are two root solvers in \mpsolve:
	the original \unisolve\ \cite{bini-fiorentino:design:00}
	which is based on the Ehrlich-Aberth iteration
	and the new hybrid algorithm called
	\secsolve\ \cite{bini-robol:secular:14}.
	These are called with the commands
	\texttt{mpsolve -au -Gi -o$\gamma$ -j1} and
	\texttt{mpsolve -as -Gi -o$\gamma$ -j1} (respectively). 
	\texttt{-Gi} means that \mpsolve\ tries to find
	for each root a unique complex disc containing it, such that 
	Newton iteration is guaranteed to converge
	quadratically toward the root starting from the center of the disc.
	\texttt{-o$\gamma$} means that
	$10^{-\gamma}$ is used as an escape bound, \emph{i.e.,} the algorithm
	stops when the complex disc containing the root has radius
	less that $10^{-\gamma}$, regardless of whether it is isolating or not.
	Unless explicitly specified, we set $\gamma=16$.
	\texttt{-j1} means that the process is not parallelized.
	Although \mpsolve\ does not do general local search,
	it has an option to search only within the unit disc.
	This option does not seem to lead to much improvement.  

\section{Improved Soft Pellet Test} 		\label{sec:pellet}
	\begin{figure}[hbt]
	\begin{progb}{
		\lline[-5] $\wtTGk(\Delta,k)$ \Comment{$f(z)$ is implicit argument}
 	    \lline[0] \OUTPUt: $r\in\{-1,0\ldots,k\}$
	    \lline[10] ASSERT: if $r\geq0$, then $\#_f(\Delta)=r$
	    \lline[5] $L\ass 53$, $d\ass \deg(f)$,
	    		$N\ass 4+\ceil{\log_2(1+\log_2(d))}$, $i\ass 0$
	    \lline[5] $\tilde{f} \ass $getApproximation( $f$, $L$ )
	    \lline[5] $\tilde{f} \ass $TaylorShift( $\tilde{f}$, $\Delta$)
	    \lline[5] While {$i\le N$}
	    \lline[10]	 Let $\tilde{f}$ be
	    		the $i$-th Graeffe iteration of $\tilde{f}$
	    \lline[10]	 $r\ass 0$
    	    \lline[10]	While {$r\leq k$}
      	    \lline[15]	$j \ass$ \intcompare( $|\tilde{f}|_r$, $\sum_{k\neq r}|\tilde{f}|_k$,$2^{-L}$)
			\footnote{$\intcompare(\tilde{a},\tilde{b},2^{-L})$ compares 
			$L$-bit approximations of real numbers $a$ and $b$.
			It returns $\true$ (resp. $\false$) only if $a>b$ (resp. $a<b$ or $a$ and $b$ are closest than a constant).
			It returns $\unresolved$ when $L$ is too small to conclude.}
	    \lline[15] While {$j = \unresolved$}
	    \lline[20] 	    $L\ass 2L$
	    \lline[20] 	    $\tilde{f} \ass$getApproximation( $f$, $L$ )
	    \lline[20] 	    $\tilde{f} \ass$TaylorShift($\tilde{f}, \Delta$ )
	    \lline[20] 	    Let $\tilde{f}$ be $i$-th Graeffe iteration of
	    			$\tilde{f}$
	    \lline[20] 	    $j \ass$\intcompare( $|\tilde{f}|_r$, $\sum_{k\neq r}|\tilde{f}|_k$,$2^{-L}$)
	    \lline[15]	 If {$j=\true$} then \Return\ $r$
	    \lline[15]	 $r\ass r+1$
	    \lline[10]	 $i\ass i+1$
	    \lline[5]	 \Return~$-1$
	    }
	\end{progb}    
	\label{fig:Tk}
	    \vspace*{-1em}
	\caption{
	\remiOK{$T_ktest(P, \Delta, k)$}{$\wtTGk(\Delta,k)$. $|\tilde{f}|_i$ is the absolute value of the coefficient
	of the monomial of degree $i$ of $\tilde{f}$, for $0\leq i \leq d$.}
	}
	\end{figure}

	The key predicate in \cite{becker+4:cluster:16} 
	is a form of Pellet test denoted
	$\wtTGk(\Delta,k)$ (with implicit $f(z)$).
	This is modified in \refFig{Tk} by adding an outer while-loop
	to control the number of Graeffe-Dandelin iterations.%
	\panOK{
	    Actually they are Dandelin's iterations.
	    See A.S. Householder. ``Dandelin, Lobachevskii, or Graeffe''
	    {\em Amer. Math. Monthly}, {\bf 66}, 464--466 (1959).
	}{}
	We try to get a definite decision (i.e., anything other than
	a \unresolved) from the soft comparison for the current Graeffe
	iteration. This is done by increasing the precision $L$ for
	approximating the coefficients of $\tilde{f}$
	in the innermost while-loop. 
	Thus we have two versions of our algorithm: (V1) uses the
	original $\wtTGk(\Delta,k)$ in \cite{becker+4:cluster:16},
	and (V2) uses the modified form in \refFig{Tk}.
	Let $\tau$V1 and $\tau$V2 be timings for the 2 versions.
	\refTab{V1V2V3} 
	shows the time $\tau$V1 (in seconds) and the ratio $\tau$V1/$\tau$V2.
	We see that (V2) achieves a consistent 2.3 to 3-fold speed up.

	\begin{table}[!h]
	\begin{center}
	    {\scriptsize
	\begin{tabular}{l||c|c||c|c||c|c||}
	    &\multicolumn{2}{c||}{ V1 }
		&\multicolumn{2}{c||}{ V2 }
		&\multicolumn{2}{c||}{ V3 }\\\hline
	    & (n1, n2, n3) & $\tau$V1 
		& (n1, n2, n3) & $\tau$V1/$\tau$V2
		& (n1, n2, n3) & $\tau$V1/$\tau$V3
	    			\\\hline\hline
	    $\Ber_{64}(z)$ & (2308,686,20223) & 19.6
		& (2308,686,6028) & 2.84
		& (2308,8,2291) & 7.06\\\hline
	    $\Mig_{64}(z;14)$ & (2060,622,18018) & 17.3
		& (2060,622,5326) & 3.03
		& (2060,20,2080) & 7.68\\\hline
	    $\Wil_{64}(z)$ & (2148,674,18053) & 23.6
		& (2148,674,5692) & 2.74
		& (2148,0,2140) & 7.23\\\hline
	    $\Spi_{64}(z)$ & (2512,728,22176) & 22.2
		& (2512,728,6596) & 2.39
		& (2512,15,2670) & 4.46\\\hline
	    $\WilM_{(11)}(z)$ & (724,202,6174) & 9.69
		& (724,202,2684) & 2.30
		& (724,18,2065) & 3.37\\\hline
	    $\MigC_{64}(z;14,3)$ & (2092,618,18515) & 20.0
		& (2092,618,5600) & 3.00
		& (2092,12,2481) &6.57\\\hline
	    $\NesC_{(4)}(z)$ & (3532,1001,30961) & 90.2
		& (3532,1001,9654) & 3.09
		& (3532,24,4588) & 6.81\\\hline
	\end{tabular}
	\vspace*{1em}
	    }
	\caption{Solving within the initial box
	    $[-50,50]^2$ with $\epsilon=2^{-53}$ with versions (V1),
	    (V2) and (V3) of \ccluster.
	n1: number of discarding tests. 
	n2: number of discarding tests returning -1 (inconclusive).
	n3: total number of Graeffe iterations.
	$\tau$V1 (resp. $\tau$V2, $\tau$V3): sequential time for
	V1 (resp. V2, V3) in seconds.
	}
	\label{tab:V1V2V3}
	\vspace*{-3em}
	\end{center}
	\end{table}
	
	In (V2), as in \cite{becker+4:cluster:16}, we use
	$\wtTGo(\Delta)$ (defined as $\wtTGk(\Delta,0)$)
	 to prove that a box $B$ has no root.
	We propose a new version (V3) that uses $\wtTGs(\Delta)$
	(defined as $\wtTGk(\Delta,d)$, where $d$ is the degree of $f$)
	instead of
	$\wtTGo(\Delta)$ to achieve this goal: instead of
	just showing that $B$ has no root, it upper bounds $\#_f(B)$.
	Although counter-intuitive, this yields a substantial improvement
	because it led to fewer Graeffe iterations overall.
	The timing for (V3) is $\tau$V3, but we display only
	the ratio $\tau$V1/$\tau$V3 in the last column
	of \refTab{V1V2V3}.  This ratio shows that (V3) enjoys
	a 3.3-7.7 fold speedup.
	Comparing $n3$ for (V2) and (V3) explains this speedup.
	\cheeOK{
	    This explanation of the counter-intuitive
	technique is still not clear -- should elaborate when
	we have space in full paper!
	}{}

\section{Filtering} 
\vspace*{-1em}
	
	A technique for speeding
	up the evaluation of predicates is the idea of filters
	(e.g., \cite{bbp:interval-filter:01}).
	The various Pellet tests can be viewed
	as a box predicate $C$ that maps
	a box $B\ib\CC$ to a value\footnote{
	    We treat two-valued predicates for simplicity;
	    the discussion could be extended to predicates
	    (like $\wtTGs$)
	    which returns a finite set of values.
	}
	in $\{\true,\false\}$.
	If $C^-$ is another box predicate with property that
	$C^-(B)=\false$ implies $C(B)=\false$, we call $C^-$ a
	\dt{falsehood filter}. 
	If $C^-$ is efficient relatively to $C$,
	and ``efficacious'' (informally, $C(B)=\false$ is likely
	to yield $C^-(B)=\false$), then it is useful to first compute
	$C^-(B)$.  If $C^-(B)=\false$, we do not need to compute $C(B)$.
	The predicate $C_0$ used in \ccluster~ is defined as follows:
	$C_0(B)$ is $\true$ if $\wtTGs(\Delta_B)$ returns $0$
	(then $B$ contains no root of $f$)
	and is $\false$ if $\wtTGs(\Delta_B)$ returns $-1$ or $k>0$
	(then $B$ may contain some roots of $f$).
	We next present the falsehood filter $C_0^-(B)$ for $C_0$.
	
	Let $f_{\Delta}$ denote the Taylor shift of $f$ in $\Delta$,
	$f_{\Delta}^{[i]}$ its $i$-th Graeffe iterate, 
	$(f_{\Delta}^{[i]})_j$ the $j$-th coefficient of $f_{\Delta}^{[i]}$,
        and $|f_{\Delta}^{[i]}|_j$ the absolute value of the $j$-th
	coefficient.  Let $d$ be the degree of $f$.
        The assertion below is a direct consequence of the 
        classical test of Pellet 
        (see \cite{becker+3:cisolate:18}[p.~12])
        and
        justify the correctness of our filters:\\
	$(A)$ if
	$|f_{\Delta}^{[N]}|_0\leq|f_{\Delta}^{[N]}|_1+|f_{\Delta}^{[N]}|_d$
              then $\wtTGs(\Delta)$ returns $-1$ or $k>0$.\\
	Our $C_0^-$ filter computes $|f_{\Delta}^{[N]}|_0$, 
	$|f_{\Delta}^{[N]}|_1$ and $|f_{\Delta}^{[N]}|_d$ and 
	checks hypothesis of $(A)$ using \intcompare.
	$|f_{\Delta}^{[N]}|_0$ and $|f_{\Delta}^{[N]}|_d$ can 
	respectively be computed as $(|f_{\Delta}|_0)^{2^N}$ and 
	$(|f_{\Delta}|_d)^{2^N}$.
	$|f_{\Delta}^{[N]}|_1$ can be computed with
	the following well known formula:
		\begin{equation}\label{eq:1}
		(f^{[i+1]}_{\Delta})_k = (-1)^k((f^{[i]}_{\Delta})_k)^2 + 
			2\sum\limits_{j=0}^{k-1}
			(-1)^j(f^{[i]}_{\Delta})_j(f^{[i]}_{\Delta})_{2k-j}
		\end{equation}
	Obtaining $|f^{[N]}_{\Delta}|_1$ with eq.~(\ref{eq:1}) requires to know
	$2^{N-1}+1$ coefficients of $f^{[1]}_{\Delta}$,
	$2^{N-2}+1$ coefficients of $f^{[2]}_{\Delta}, \ldots,$
	and finally $3=2^1+1$ coefficients of $f^{[N-1]}_{\Delta}$.
	In particular, it requires to compute entirely the iterations
	$f^{[i]}_{\Delta}$ such that $2^{N-i}\leq d$, and it is 
	possible to do it more efficiently that with eq.~(\ref{eq:1})
	(for instance with the formula given in definition 2 of 
	\cite{becker+3:cisolate:18}).

Our $C_0^-$ filter takes as input a precision $L$, the Taylor shift 
$f_{\Delta}$ of the $L$ bit approximation of $f$
and its $i$-th Graeffe iteration $f^{[i]}_{\Delta}$
such that $2^{N-i}\leq \frac{d}{4}$ and $2^{N-(i+1)}> \frac{d}{4}$.
It computes $|f_{\Delta}^{[N]}|_0$, $|f_{\Delta}^{[N]}|_d$
and the $2^{N-j}+1$ first coefficients of $f^{[j]}_{\Delta}$ 
for $i<j\leq N$ with eq.~(\ref{eq:1}).
Then it checks the hypothesis of $(A)$ using \intcompare,
and returns $\false$ if it is verified, and $\true$ otherwise.
In practice, it is implemented within the procedure implementing 
$\wtTGs(\Delta_B)$.

Incorporating $C_0^-$ into Version (V3), we obtain (V4)
	and the speed up can be seen in \refTab{V3V4}.  
	Filtering with $C_0^-$ becomes more effective as degree grows
	and this is because one has $2^{N-i}\leq \frac{d}{4}$ for smaller 
	$i$ (recall that $N=4+\ceil{\log_2(1+\log_2(d))}$).

	\begin{table}[!h]
	\vspace*{-2em}
	    {\scriptsize
	\begin{center}
	\begin{tabular}{l|l||c|c||c|c||}
	    \multicolumn{2}{c||}{}
	    &\multicolumn{2}{c||}{ V3 }&\multicolumn{2}{c||}{ V4 }\\\hline
	    & & n3 & $\tau$V3 & n3 & $\tau$V3/$\tau$V4\\\hline\hline
	    \multirow{4}{*}{$\Ber_d(z)$}
			& $d=64$  & 2291 & 2.61& 2084  & 1.08\\
			& $d=128$ & 4496 & 14.5& 3983  & 1.13\\
			& $d=256$ & 8847 & 94.5& 7714  & 1.19\\
			& $d=512$ & 15983 & 620& 11664 & 1.42\\
			& $d=767$ & 19804 & 1832& 13863 & 1.53\\\hline
	    \multirow{4}{*}{$\Mig_d(z;a)$}
			& $(d,a)=(64,14)$  & 2080 & 2.41& 1808 & 1.22\\
			& $(d,a)=(128,14)$ & 3899 & 12.1& 3257 & 1.21\\
			& $(d,a)=(256,14)$ & 7605 & 88.3& 6339 & 1.33\\
			& $(d,a)=(512,14)$ & 15227& 674& 10405 & 1.57\\\hline
	    \multirow{4}{*}{$\Wil_d(z)$}
			& $d=64$ & 2140 & 3.27& 1958  & 1.05\\
			& $d=128$ & 2240 & 10.0& 1942 & 1.09\\
			& $d=256$ & 2414 & 36.6& 2108 & 1.21\\
			& $d=512$ & 2557 & 129& 1841  & 1.43\\\hline
	    \multirow{4}{*}{$\Spi_d(z)$}
			& $d=64$ & 2670 & 4.43& 2364    & 1.08\\
			& $d=128$ & 5090 & 28.8& 4405   & 1.07\\
			& $d=256$ & 9746 & 182& 8529    & 1.10\\
			& $d=512$ & 19159 & 1340& 14786 & 1.19\\\hline
	    \multirow{4}{*}{$\WilM_{(D)}(z)$}
			& $(D,d)=(11,66)$  & 2065 & 2.87& 1818 & 1.14\\
			& $(D,d)=(12,78)$  & 2313 & 3.95& 2053 & 1.12\\
			& $(D,d)=(13,91)$  & 2649 & 5.89& 2336 & 1.18\\
			& $(D,d)=(14,105)$  & 2892& 8.56& 2537 & 1.29\\\hline
	    \multirow{4}{*}{$\MigC_{d}(z;a,k)$}
		& $(d,a,k)=(64,14,3)$ & 2481 & 2.94& 2145  & 1.13\\
		& $(d,a,k)=(128,14,3)$ & 4166 & 14.4& 3555 & 1.16\\
		& $(d,a,k)=(256,14,3)$ & 7658 & 86.0& 6523 & 1.27\\
		& $(d,a,k)=(512,14,3)$ & 15044& 650& 10472 & 1.63\\\hline
	    \multirow{4}{*}{$\NesC_{(D)}(z)$}
			& $(D,d)=(4,27)$ & 1628 & 0.77& 1459   & 1.07\\
			& $(D,d)=(5,81)$ & 4588 & 13.2& 4085   & 1.12\\
			& $(D,d)=(6,243)$ & 13056 & 358& 11824 & 1.26\\\hline
	    \end{tabular}
	\end{center}
	    }
		\caption{Solving within the initial box $[-50,50]^2$
	with $\epsilon=2^{-53}$ with versions (V3), (V4) of \ccluster.
	n3: number of Graeffe iterations.
	$\tau$V3 and $\tau$V4: sequential time in seconds.
	    }
	\label{tab:V3V4}
	\vspace*{-2em}
	\end{table}

\section{Escape Bound}		\label{sec:escape}
	The $\epsilon$ parameter is usually understood as the precision
	desired for roots.  But we can also view it as an escape bound
	for multiple roots as follows: we do not refine a disc
	that contains a simple root, even if its radius is $\ge \epsilon$. 
	But for clusters of size greater than one,
	we only stop when the radius is $<\epsilon$.
	\mpsolve\ has a similar option.
	This variant of (V4) is denoted (V4'). We see 
	from \refTab{V4escape} that (V4') gives a modest improvement
	(up to 25\% speedup) over (V4) when $-\log\epsilon=53$.
	This improvement generally grows with $-\log\epsilon$
	(but $\WilM_{(11)}(z)$ shows no difference).
	\begin{table}[!h]
	\begin{center}
	   {\scriptsize
	\begin{tabular}{l||c|c|c||c|c|c||}
	    &\multicolumn{3}{c||}{(V4)}&\multicolumn{3}{c||}{(V4')}\\\hline
	    \hfill{}$\epsilon$:\hspace*{2mm}
	    \rule[-.3\baselineskip]{0pt}{4mm}
	   		& $2^{-53}$ & $2^{-530}$ & $2^{-5300}$
	    		& $2^{-53}$ & $2^{-530}$ & $2^{-5300}$\\\hline
	    &$\tau$53 (s)& $\tau$530/$\tau$53 & $\tau$5300/$\tau$53
		& $\tau$53 (s)& $\tau$530/$\tau$53 & $\tau$5300/$\tau$53
	    		\\\hline\hline
	$\Ber_{64}(z)$		& 2.42& 1.26& 4.22& 1.99& 0.94& 0.94\\\hline
	$\Mig_{64}(z;14)$	& 1.97& 1.63& 4.56& 1.61& 1.45& 1.38\\\hline
	$\Wil_{64}(z)$ 		& 3.22& 1.10& 2.16& 2.91& 0.96& 1.01\\\hline
	$\Spi_{64}(z)$		& 4.09& 1.33& 5.25& 3.05& 0.95& 0.95\\\hline
	$\WilM_{(11)}(z)$	& \cored{2.51}& \cored{1.12} & \cored{2.03}
	    		& \cored{2.50}& \cored{1.13}& \cored{1.98}\\\hline
	$\MigC_{64}(z;14,3)$	& 2.60& 1.89& 4.15& 2.20& 1.70& 1.80\\\hline
	$\NesC_{4}(z)$		& 11.9& 1.08& 2.67& 10.4& 1.00& 0.99\\\hline
	\end{tabular}
	\vspace*{1em}
	    }
	\caption{Solving within the box $[-50,50]^2$
	with versions (V4) and (V4') of \ccluster~ with three values of
	    $\epsilon$. 
	$\tau$53 (resp. $\tau$530, $\tau$5300):
	    sequential time for (V4) and (V4') in seconds.
	}
	\label{tab:V4escape}
	    \vspace*{-2em}
	\end{center}
	\end{table}

\section{Conclusion}
	Implementing subdivision algorithms is relatively easy
	but achieving state-of-art performance requires
	much optimization and low-level development.
	This paper explores several such techniques.
	We do well compared to \fsolve\ in \maple, but
	the performance of \mpsolve\ is superior
	to the global version of \ccluster. But \ccluster\ can still shine
	when looking for local roots or when $\vareps$ is large.

\bibliographystyle{abbrv}
\bibliography{references}

\end{document}